\documentclass[twocolumn,aps,superscriptaddress,prl,showpacs]{revtex4-1}

\usepackage{graphicx}% Include figure files
\usepackage{dcolumn}% Align table columns on decimal point
\usepackage{bm}% bold math

\begin{document}

\title{Observation of Intense Second Harmonic Generation from MoS$_{2}$ Atomic Crystals}

\author{Leandro M. Malard}
\affiliation{Department of Physics, Federal University of Minas Gerais, 30123-970 Belo Horizonte-MG, Brazil}
\author{Thonimar V. Alencar}
\affiliation{Department of Physics, Federal University of Minas Gerais, 30123-970 Belo Horizonte-MG, Brazil}
\author{Ana Paula M. Barboza}
\affiliation{Department of Physics, Federal University of Minas Gerais, 30123-970 Belo Horizonte-MG, Brazil}
\author{Kin Fai Mak}
\affiliation{Department of Physics, Columbia University, 10027 New York, USA}
\author{Ana M. de Paula}
\email{ana@fisica.ufmg.br}
\affiliation{Department of Physics, Federal University of Minas Gerais, 30123-970 Belo Horizonte-MG, Brazil}

\begin{abstract}
The nonlinear optical properties of few-layer MoS$_2$ two-dimensional crystals are studied 
using femtosecond laser pulses. We observed highly efficient second harmonic generation from 
the odd-layer crystals, which shows a polarization intensity dependence that directly reveals
the underlying symmetry and orientation of the crystal.
Additionally, the measured second-order susceptibility spectra provide information about the
electronic structure of the material. Our results open up new opportunities for studying the
non-linear optical properties in these novel 2D crystals.

\end{abstract}


\maketitle

The family of 2-dimensional (2D) transition metal dichalcogenide semiconductors MX$_2$ 
(M for Mo or W, and X for S, Se or Te) has attracted much recent attention. 
They are direct band gap semiconductors that possess properties very different from 
graphene~\cite{novoselov2005,novoselov2011,coleman2011,novoselov2012,mak2010,splendiani2010,mak2012,zeng2012,cao2012,tongay2012,wang2012}. 
For instance, single layer MoS$_2$ shows efficient light emission~\cite{mak2010,splendiani2010}, 
optical control of valley polarization~\cite{mak2012,zeng2012,cao2012}, enhanced many-body
interactions~\cite{mak2012b}, and transistors with high on-off ratios~\cite{radisavljevic2011,yin2012}. 
These properties are not only important for applications in electronics and optoelectronics 
with 2D crystals, the associated valley-dependent physics and topological transport 
phenomena~\cite{xiao2012} are also of fundamental interests.

The lack on an inversion center in the crystalline structure of the material (so that energy 
gaps at the K and K$^{'}$ valleys with finite but opposite Berry curvatures are introduced) 
is the underlying reason for developing interesting properties. Actually, the inversion
asymmetry of the material not just affects its electronic and linear optical properties, but
also gives rise to a finite second-order optical nonlinearity ($\chi^{(2)}\neq 0$)~\cite{boyd,shen}. 
The finite $\chi^{(2)}$ provides opportunities for new applications in
optoelectronics devices (e.g. coherent control of valley-polarized currents) that are not 
possible with graphene. 

In this Letter we report the observation of intense second harmonic generation (SHG) from 
odd-layered MoS$_2$ thin films. 
In contrast, very low SHG is observed for samples with even-layer 
numbers due to the restoration of inversion symmetry. Moreover, the 3-fold rotation symmetry 
of the crystal produces a characteristic 6-fold polarization dependence pattern for the SHG 
intensity, from which we have developed an optical method for imaging the underlying crystalline
structure and domain orientations with sub-micron resolution. 
We have also measured the $\chi^{(2)}$ spectra by varying the incident pump photon energy.
Comparison to absorption measurements shows that the observed efficient SHG comes from the
resonantly enhanced virtual optical transitions in the near UV range.

The MoS$_{2}$ mono and few-layers were obtained by mechanical exfoliation of natural 
bulk hexagonal MoS$_{2}$ which is deposit in Si covered with 300~nm SiO$_{2}$ 
substrates~\cite{novoselov2005} or on transparent amorphous quartz (SPI Supplies Inc.). 
We have characterized the MoS$_{2}$ layer number with a combination of optical microscopy, 
AFM~\cite{lee2010}, Raman spectroscopy \cite{lee2010} and photoluminescence imaging~\cite{mak2010}. 
For the second harmonic imaging we used a 140~fs Ti-Sapphire oscillator (Coherent Chameleon) 
with 80~MHz repetition and tunable  wavelength from 680~nm to 1080~nm which is directed to 
a confocal scanning laser microscope (Olympus FV300) modified for two photon excitation. 
The laser is focused on the sample at normal incidence by a 60$\times$ objective (NA 0.75).
The back reflected signal is then directed to a dichroic mirror and a thin band pass centered 
at the SH wavelength to completely remove the laser scattered light and the SH signal 
is detect by a photomultiplier tube. 

Fig~\ref{sample}(a) shows the optical images of mono- and few-layer MoS$_2$ samples,
where the colors indicate the number of layers~\cite{benameur2011}. 
Fig.~\ref{sample}(b) shows the sample characterization by atomic force microscopy (AFM) at 
the dashed triangle area marked in Fig.~\ref{sample}(a).  
By measuring the height of MoS$_{2}$ relative to the substrate~\cite{lee2010} we found regions 
of MoS$_{2}$ monolayer (1L), bilayer (2L) and trilayer (3L) respectively, as indicated by 
the labels at Fig.~\ref{sample}(b), see supplemental materials for details. 
Fig.~\ref{sample}(c) presents the sample second harmonic image, same area as in fig.~\ref{sample}(a), showing well defined regions where the second harmonic intensity varies. 
To unambiguously show that the strong optical emission in Fig.~\ref{sample}(c) is due to 
the optical second harmonic generation, we measured its intensity 
dependence on pump laser power as shown in Fig.~\ref{sample}(d). Since SHG is a
second order nonlinear optical process, the SHG intensity scales quadratically with the
fundamental pump intensity~\cite{boyd,shen}. 
This is in agreement with our results shown in Fig.~\ref{sample}(d) where the log scale plot 
shows a linear dependence with a slope equals to $2.00\pm0.02$.

\begingroup
\begin{figure*}[ht]
\centering
\includegraphics [scale=0.45]{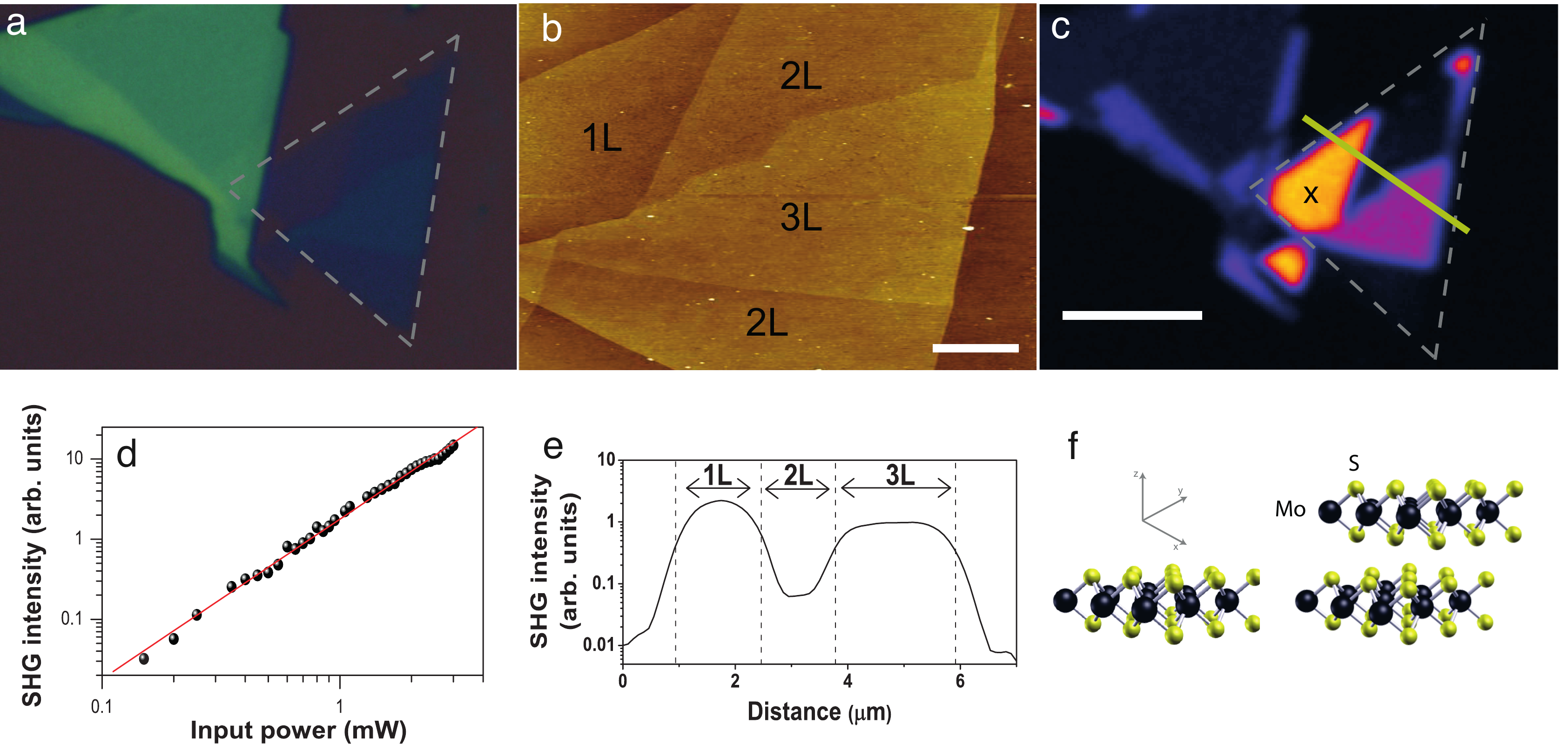}
\caption{\label{sample} (color online) (a) Optical microscope image of the MoS$_2$ thin film. 
(b) AFM image of the same MoS$_2$ thin film made within the dashed triangle shown in
part a, scale bar 1~$\mu$m. By measuring the height relative to the
substrate, the number of MoS$_2$ layers are determined as 1L for monolayer, 2L
for bilayer and 3L for trilayer MoS$_2$. (c) Second Harmonic image
collected from the same MoS$_2$ thin film, pump laser wavelength at 800~nm,
1.55~eV. Brighter colors means stronger SHG intensity, and the scale bar is
5~$\mu$m. (d) Log vs. Log plot of the power dependence of SHG from the X
marked position at part c. The circles are the measured data and the red
line is a linear fit of the data with extracted slope of $2.00\pm0.02$. 
(e) Intensity profile of the SHG image from left to right at the yellow line shown
in part c. (f) Atomic structure of monolayer (left) and bilayer
(right) MoS$_2$.}
\end{figure*}
\endgroup

To visualize the spatial variation of the second harmonic intensity as a
function of the MoS$_{2}$ ultra-thin film position we plot the SHG intensity
profile in log scale, Fig.~\ref{sample}(e) at the marked yellow line in Fig.~\ref{sample}(c). 
It is intriguing to observe that a stronger signal is present for the monolayer
MoS$_{2}$ when compared to the trilayer MoS$_{2}$, for the pump wavelength at
800~nm, considering that less material is present in the former case. Also the
bilayer MoS$_{2}$ shows very low second harmonic generation, almost two orders 
of magnitude lower than the monolayer. 
The small observed signal may come from the boundary between the layers~\cite{tom1983,shen89}.
The $\chi^{(2)}$ tensor holds the information of the crystallographic properties of the 
crystal~\cite{boyd,shen} and for SHG without phase matching determines the SHG intensity.
This is true for the present case where the sample thickness is much smaller than the light wavelength.
Hence, in order to understand this spatial intensity
variation, we should determine each element of the tensor to directly probe the
crystal structure of two-dimensional MoS$_{2}$. 

The bulk MoS$_{2}$ crystal has a trigonal prismatic structure with Bernal
stacking~\cite{mattheiss1973}, thus it belongs to the P6$_{3}$/mmc non-symmorphic
space group~\cite{mattheiss1973,dresselhaus}, with an inversion
symmetry operation in the middle of the two MoS$_{2}$ monolayers. 
The monolayer and bilayer MoS$_{2}$ structures are shown at Fig.~\ref{sample}(f).
As the layer number decreases down to one monolayer of MoS$_{2}$, it is
convenient to make the point group representation of the unit cell, which is
comprised by three atoms (one Mo and two S). Therefore we assign the symmetry of
the monolayer as D$_{3h}$, since the inversion symmetry present in bulk
MoS$_{2}$ is lost. However, in bilayer MoS$_{2}$ the inversion symmetry
operation is again present, nonetheless due to the lack of translation symmetry
along the z-axis, the point group is D$_{3d}$. 
In general, even number of layers belongs to D$_{3d}$ point group and odd number of layers to the D$_{3h}$ point group (similar to few-layers graphene point groups~\cite{malard2009}). 
This simple approach can explain qualitatively the absence of second harmonic generation in
bilayer MoS$_{2}$: due to the presence of inversion symmetry the $\chi^{(2)}$
tensor is equal to zero, thus no second harmonic is generated for bilayer
MoS$_{2}$ within the dipole approximation~\cite{boyd,shen}. Generally speaking,
there is no second harmonic generation for even MoS$_{2}$ layer number, but it
can be generated in odd number of MoS$_{2}$ layers. 
We should note that, we also studied monolayer graphene exfoliated on transparent amorphous 
quartz and observed no significant SHG signal, as is expected from the symmetry of the crystal.

Due to the sensitivity on crystal symmetry, polarization-resolved SHG measurements provide important crystallographic information of MoS$_2$ atomic layers.
To measure the polarization dependence of the SHG, we have placed the sample on top of a 
precision rotation stage, which is placed under the microscope objective. 
The laser (at 800~nm) is focused on the sample at normal incidence and with a fixed linear polarization that is in the plane of the sample. The sample is initially with the crystallographic axis at an arbitrary angle to the incident polarization.
We have measured one SH image for each sample angle, at 5 degree steps, 
for the SH polarization parallel to that of the incident laser (a polarization analyzer is placed before the detector to select the SH polarization). 

Figure~\ref{pol-shg}(a) shows the experimental results of the SHG polarization
dependence for the monolayer of MoS$_{2}$ in Fig.~\ref{sample}(c), 
where the second harmonic intensity is plotted as a function of the sample rotation angle. 
A clear 6-fold pattern can be observed at Fig.~\ref{pol-shg}(a).
In figure~\ref{pol-shg}(b) is a sketch of a top view showing the angle of the
MoS$_{2}$ crystallographic orientation to the incident laser polarization
direction ($\hat{e}_{\omega}$).
Figure.~\ref{pol-shg}(c) shows the SH images for the monolayer and trilayer
MoS$_2$ samples together with the crystallographic orientation of the crystal
lattice obtained from the polarization data in Fig.~\ref{pol-shg}(a). 
To obtain the crystallographic orientation, we first describe the electric field of 
the generated second harmonic light ({\bf E}(2$\omega$)) along a given direction 
($\hat{e}_{2\omega}$) in terms of the $\chi^{(2)}$ tensor and input light polarization 
vector ($\hat{e}_{\omega}$) as~\cite{boyd,shen}:

\begin{equation}\label{eq1}
\mathbf{E}(2\omega)\cdot\hat{e}_{2\omega}=C\hat{e}_{2\omega}\cdot\mathbf{\chi^{
(2)}}:\hat{e}_{\omega}\hat{e}_{\omega}
\end{equation}
where $\omega$ is the laser frequency, $2\omega$ is the SH frequency and $C$ is
a proportionality constant which contains local field factors determined by the
local dielectric environment. Equation~\ref{eq1} fully describes the
polarization dependence of the SHG, and its specific form will depend on the
$\chi^{(2)}$ tensor. For the odd-layer with D$_{3h}$ point group symmetry, the
second order susceptibility tensor has a single non zero
element~\cite{boyd,shen}:
$\chi^{(2)}_{MoS_{2}}$$\equiv$$\chi^{(2)}_{xxx}$=-$\chi^{(2)}_{xyy}$=-$\chi^{(2)}_{yyx}$=-$\chi^{(2)}_{yxy}$,
where $x$ corresponds to the armchair direction, since it has a mirror plane
symmetry~\cite{boyd,shen}, and $y$ is the zigzag direction, see
Fig.~\ref{pol-shg}b. Thus, using Eq.~\ref{eq1} 
we obtain the resulting dependence of the generated second harmonic electric
field as a function of the sample angle for a pump laser polarization
($\hat{e}_{\omega}$) parallel to the analyzer ($\hat{e}_{2\omega}$), that can be
expressed as~\cite{boyd,shen}:

\begin{equation}\label{eq2}
E(2\omega)=C\chi^{(2)}_{MoS_{2}}cos(3\phi+\phi_0)
\end{equation}
where $\phi$ is the angle between input the laser polarization and the $x$ direction, and
$\phi_0$ is the initial crystallographic orientation of MoS$_{2}$ sample. 
It is worth mentioning that for the analyzer at crossed polarization with the input laser the second harmonic electric field is proportional to $sin(3\phi+\phi_0)$, hence the total SH intensity (without analyzer) is constant with $\theta$.
The intensity of the generated second harmonic light as a function of
sample angle is obtained from Eq.~\ref{eq2} \cite{boyd,shen}, resulting in $I_{2\omega}\propto cos^{2}(3\phi +\phi_{0})$. Using this intensity dependence of the second harmonic with sample angle, we have fitted the experimental data as shown in Fig.~\ref{pol-shg}(a). The fitting
shown by the red lines agrees with the experimental data within $\phi_0=16\pm5$
degrees, indicating that the armchair ($x$) of MoS$_{2}$ was initially rotated
with respect to input laser polarization. We should note that as our measurement is phase insensitive, there is an arbitrariness of 60 degrees in the definition of the $x$-axis.

\begingroup
\begin{figure}[ht]
\centering
\includegraphics [scale=0.3]{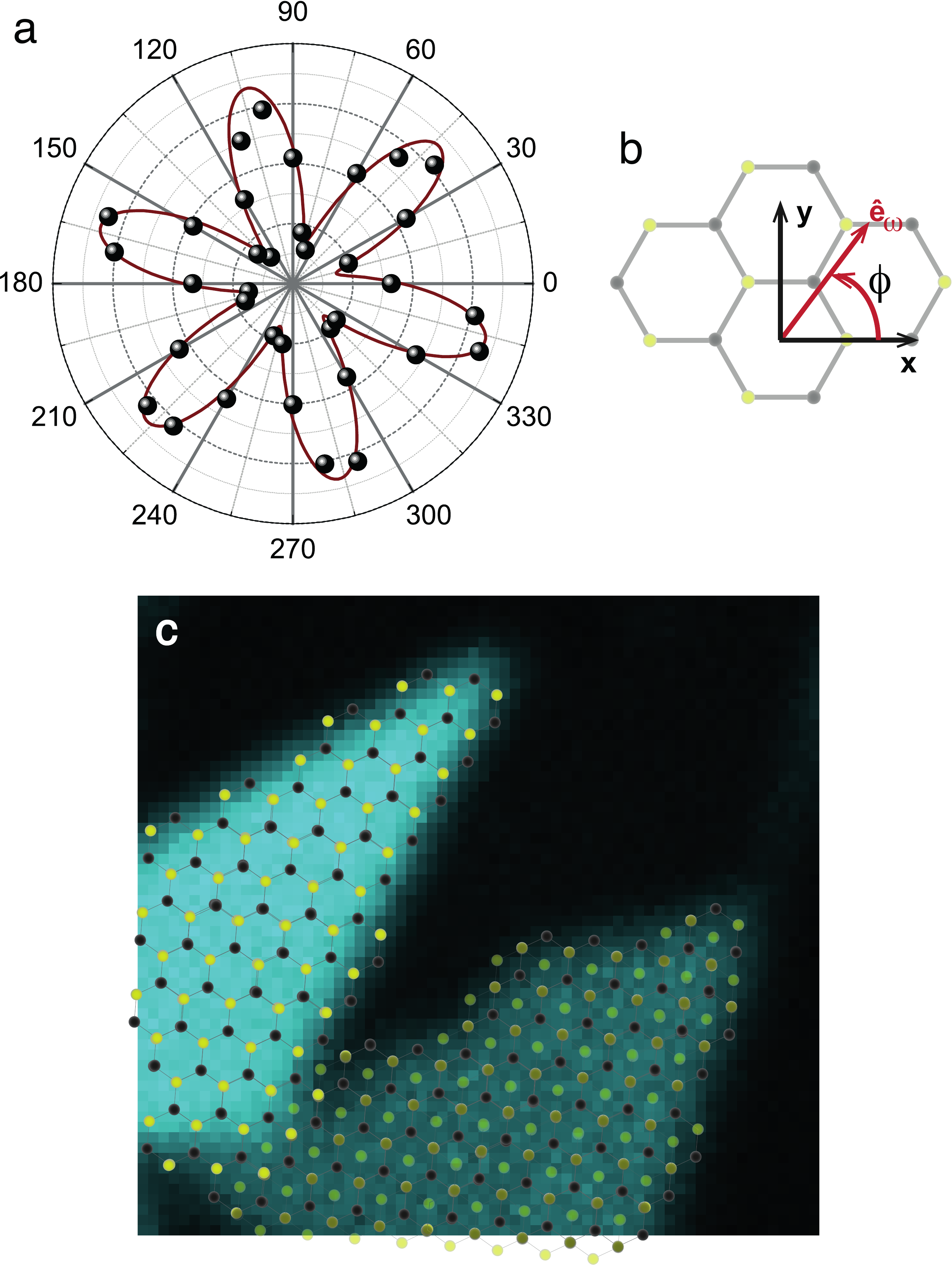}
\caption{\label{pol-shg} (color online) (a) Polar plot of the second harmonic intensity from monolayer
MoS$_{2}$ as a function of the sample angle. (b) Top view of the MoS$_{2}$
crystallographic orientation with respect to the incident laser polarization
($\hat{e}_{\omega}$). (c) SH image of the same sample as in
Fig.~\ref{sample}c showing the crystallographic direction of the monolayer and
trilayer MoS$_2$ sample determined by our polarization measurements.}
\end{figure}
\endgroup

This result shows that from this simple approach, one can measure the
crystallographic direction of MoS$_{2}$ purely by optical means. As an example
of the application of this technique, we have shown in Fig.~\ref{pol-shg}(c) the
SH image of the monolayer and trilayer MoS$_2$ sample together with the
crystallographic orientation of the crystal lattice determined by the
polarization measurement in Fig.~\ref{pol-shg}(a). 
This method can be used for future applications in materials where it is important to determine the crystallographic directions.
Two well known examples are the 2D heterostructures (i.e. stacks of different 2D crystals) and nanoribbons~\cite{louie2006,jeroen2007,leroy2012} where the physical properties depends on the relative 2D crystal stacking orientation and the edge atomic structure, respectively. 

The SHG can provide also valuable information about the electronic structure of
crystals. Hence, we have measured SHG intensity as a function of the incident
laser wavelength for the monolayer and trilayer.
For these measurements the laser beam was redirected to a 
home-built microscope with 60$\times$ objective, 0.75~NA, giving a spot size of 1.2~$\mu$m in 
diameter at the sample focal plane. 
The back scattered light is collected and sent to a spectrometer and detected by a charge 
coupled device (CCD) with a set of filters to remove the pump laser light. 
In order to avoid any spectral dependence of the SHG signal coming from our experimental setup, 
we have used the same laser power and we have verified that the spot size does not change 
significantly for the different laser lines used. 
Also, the spectral dependence of the spectrometer, collection optics and CCD camera were 
normalized by using a calibration lamp (Ocean Optics). 

Figure~\ref{shg-energy}(a) shows an optical image of the MoS$_2$ sample on transparent amorphous 
quartz where we have characterized the mono-, bi- and tri-layers by AFM. 
Figures~\ref{shg-energy}(b)-(d) are SH images of such sample taken under different
incident photon energies where is possible to observe that by decreasing the
laser energy the monolayer and the trilayer SH intensity becomes very similar,
Fig.~\ref{shg-energy}(c), and at Fig.~\ref{shg-energy}(d) the monolayer intensity
becomes slightly smaller as compared to the trilayer. Figure~\ref{shg-energy}(e)
shows the SH signal for the monolayer and trilayer samples as a
function of the SH energy, where resonance enhancement peaks are clearly
observed. 
To extract this spectral dependence we
normalized the SHG intensity from MoS$_2$ ($I_{MoS_2}$) to that from
alpha-quartz ($I_{quartz}$). For samples on transparent amorphous quartz
substrates with bulk inversion symmetry, the susceptibility ratio can be written as~\cite{tom1983,shen89}
(details in supplemental materials): 

\begin{equation}\label{eq3}
\frac{\chi^{(2)}_{MoS_2}}{\chi^{(2)}_{quartz}} \approx \frac{\pi (N.A.)^2
f}{4nk_0Nd}\frac{\sqrt{I_{MoS_2}(2\omega)}/I_{MoS_2}(\omega)}{\sqrt{I_{quartz}
(2\omega)}/I_{quartz}(\omega)}
\end{equation}
here $n$ is the refractive index of quartz (note that we have neglected the
slight spectral variation of $n$ and the small difference of $n$ between
amorphous and crystalline quartz), $f$ is a numerical constant that depends on
the numerical aperture of the objective $(N.A.)$, $N$ is the number of layers, 
$d$ is the MoS$_{2}$ interplanar distance (0.67~nm), $k_0$=$2\pi/\lambda$ 
is the magnitude of the vacuum fundamental wavevector and
$\chi^{(2)}_{quartz}$ is the bulk susceptibility of alpha-quartz. 
Fot the bulk quartz, we have used the expression for back reflected SHG intensity
from the alpha-quartz surface~\cite{bloembergen1962}.

The $\chi^{(2)}_{MoS_2}$$\slash$$\chi^{(2)}_{quartz}$ for both the monolayer 
and trilayer samples are plotted in figure~\ref{shg-energy}(e) as a function of the SH energy, frequency $2\omega$. 
Also shown is the normalized linear absorption spectra of the corresponding samples (details of measurement setup in the Supplemental Materials). 
We can see that over the specified spectral range,
there are three peaks known as the A, B and C excitons. They originate from the
direct optical transitions at the K (for the A and B features) and the $\Gamma$
(for the C feature) point of the Brillouin zone~\cite{mak2010}. A significant
enhancement of $\chi^{(2)}_{MoS_2}$ is observed with spectral positions
very well matched with the C-resonances in the linear absorption spectra ($\sim
2.8$~eV for the monolayer and $\sim 2.7$~eV for the trilayer sample). 
It suggests that the resonance enhancement of $\chi^{(2)}_{MoS_2}$ is due to the
increased density of electronic states at the $\Gamma$ point, see inset of
Fig.~\ref{shg-energy}(e). The different resonance position for the monolayer
and the trilayers is thus a consequence of the different energy gaps at the
$\Gamma$ point of the Brillouin zone, that is a modification of the electronic structure
from quantum confinement at reduced sample thickness. The SHG from the
trilayer $\chi^{(2)}_{MoS_2}$ is consistently smaller than that of the
monolayer. 
Away from the resonances, the $\chi^{(2)}_{MoS_2}$ of the two odd layers become similar.
Such an observation demonstrates the perfect cancellation of second harmonic
(SH) dipole moments in an AB-stacked Bernal unit cell.

\begingroup
\begin{figure}[t]
\centering
\includegraphics [scale=0.9]{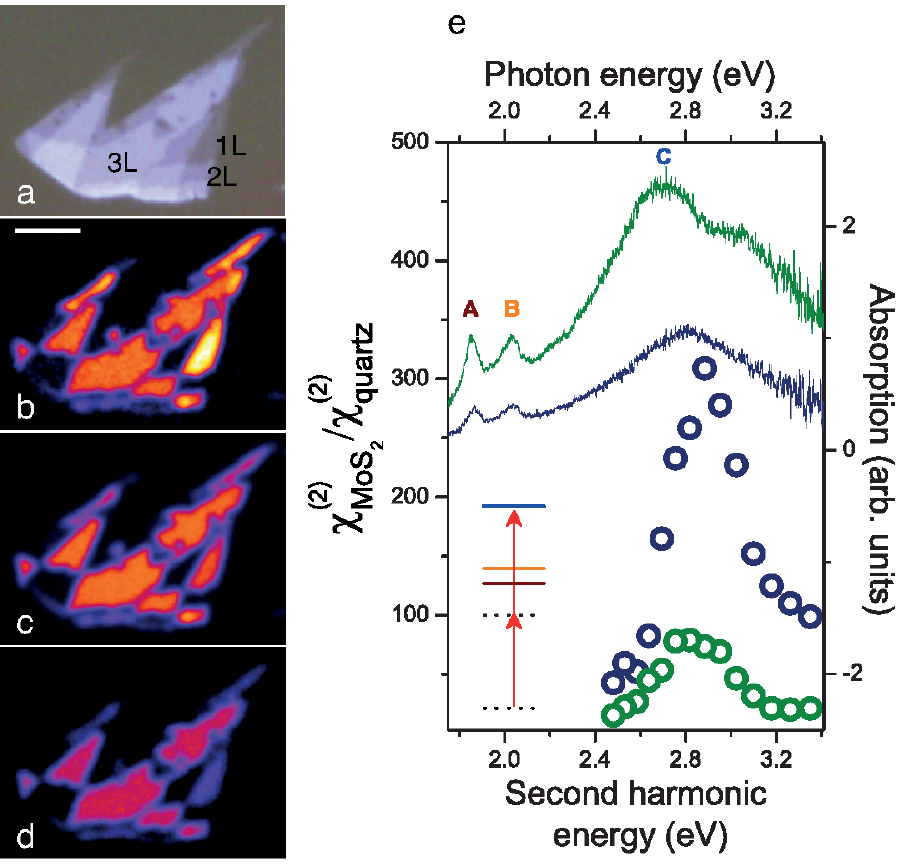}
\caption{\label{shg-energy} (color online) (a) Optical image of the few-layer MoS$_2$ sample on
transparent quartz substrate. The number of layers are indicated as measured by
AFM. (b) SH image for the same sample with pump laser wavelength at 860~nm, 1.44~eV, 
(c) 900~nm, 1.38~eV, and (d) 960~nm, 1.29~eV.  (e) Second order susceptibility,
left scale, of a sheet of MoS$_2$ for monolayer (blue circles) and trilayer
(green circles) samples with respect to the bulk quartz as a function of pump
laser energy. Measured linear absorption (right scale) spectra of monolayer
(blue line) and trilayer (green line), with the labeled optical transitions. The
inset shows the diagram of the second harmonic enhancement in monolayer MoS$_2$
where the two-photon energy is in resonance with the C absorption peak.}
\end{figure}
\endgroup

In summary, our demonstration of strong SHG from monolayer and trilayer
MoS$_{2}$, although being only a few atoms thick, shows its potential use
for nonlinear optical device applications. Also the SHG dependence on the
crystallographic axis and electronic resonance effects open up an optical way 
to characterize the crystal structure of the many novel 2D crystals. 
Furthermore, the ability to
distinguish different crystallographic axis can be applied to improve materials
properties, such as stacking different 2D materials with atomic registry and
edge-defined nanoribbons.

This work was supported by CNPq, FAPEMIG and INCT Carbon Nanomaterials-MCT, Brazil.
The authors thank B.R.A. Neves for assistance with the AFM measurements and 
T. F. Heinz, A. Jorio and G. Weber for fruitful discussions.

%\bibliography{shg}

\begin{thebibliography}{28}%
\makeatletter
\providecommand \@ifxundefined [1]{%
 \@ifx{#1\undefined}
}%
\providecommand \@ifnum [1]{%
 \ifnum #1\expandafter \@firstoftwo
 \else \expandafter \@secondoftwo
 \fi
}%
\providecommand \@ifx [1]{%
 \ifx #1\expandafter \@firstoftwo
 \else \expandafter \@secondoftwo
 \fi
}%
\providecommand \natexlab [1]{#1}%
\providecommand \enquote  [1]{``#1''}%
\providecommand \bibnamefont  [1]{#1}%
\providecommand \bibfnamefont [1]{#1}%
\providecommand \citenamefont [1]{#1}%
\providecommand \href@noop [0]{\@secondoftwo}%
\providecommand \href [0]{\begingroup \@sanitize@url \@href}%
\providecommand \@href[1]{\@@startlink{#1}\@@href}%
\providecommand \@@href[1]{\endgroup#1\@@endlink}%
\providecommand \@sanitize@url [0]{\catcode `\\12\catcode `\$12\catcode
  `\&12\catcode `\#12\catcode `\^12\catcode `\_12\catcode `\%12\relax}%
\providecommand \@@startlink[1]{}%
\providecommand \@@endlink[0]{}%
\providecommand \url  [0]{\begingroup\@sanitize@url \@url }%
\providecommand \@url [1]{\endgroup\@href {#1}{\urlprefix }}%
\providecommand \urlprefix  [0]{URL }%
\providecommand \Eprint [0]{\href }%
\providecommand \doibase [0]{http://dx.doi.org/}%
\providecommand \selectlanguage [0]{\@gobble}%
\providecommand \bibinfo  [0]{\@secondoftwo}%
\providecommand \bibfield  [0]{\@secondoftwo}%
\providecommand \translation [1]{[#1]}%
\providecommand \BibitemOpen [0]{}%
\providecommand \bibitemStop [0]{}%
\providecommand \bibitemNoStop [0]{.\EOS\space}%
\providecommand \EOS [0]{\spacefactor3000\relax}%
\providecommand \BibitemShut  [1]{\csname bibitem#1\endcsname}%
\let\auto@bib@innerbib\@empty
%</preamble>
\bibitem [{\citenamefont {Novoselov}\ \emph {et~al.}(2005)\citenamefont
  {Novoselov}, \citenamefont {Jiang}, \citenamefont {Schedin}, \citenamefont
  {Booth}, \citenamefont {Khotkevich}, \citenamefont {Morozov},\ and\
  \citenamefont {Geim}}]{novoselov2005}%
  \BibitemOpen
  \bibfield  {author} {\bibinfo {author} {\bibfnamefont {K.~S.}\ \bibnamefont
  {Novoselov}}, \bibinfo {author} {\bibfnamefont {D.}~\bibnamefont {Jiang}},
  \bibinfo {author} {\bibfnamefont {F.}~\bibnamefont {Schedin}}, \bibinfo
  {author} {\bibfnamefont {T.~J.}\ \bibnamefont {Booth}}, \bibinfo {author}
  {\bibfnamefont {V.~V.}\ \bibnamefont {Khotkevich}}, \bibinfo {author}
  {\bibfnamefont {S.~V.}\ \bibnamefont {Morozov}}, \ and\ \bibinfo {author}
  {\bibfnamefont {A.~K.}\ \bibnamefont {Geim}},\ }\href {\doibase
  10.1073/pnas.0502848102} {\bibfield  {journal} {\bibinfo  {journal} {PNAS}\
  }\textbf {\bibinfo {volume} {102}},\ \bibinfo {pages} {10451} (\bibinfo
  {year} {2005})}\BibitemShut {NoStop}%
\bibitem [{\citenamefont {Novoselov}(2011)}]{novoselov2011}%
  \BibitemOpen
  \bibfield  {author} {\bibinfo {author} {\bibfnamefont {K.~S.}\ \bibnamefont
  {Novoselov}},\ }\href {\doibase 10.1103/RevModPhys.83.837} {\bibfield
  {journal} {\bibinfo  {journal} {Rev. Mod. Phys.}\ }\textbf {\bibinfo {volume}
  {83}},\ \bibinfo {pages} {837} (\bibinfo {year} {2011})}\BibitemShut
  {NoStop}%
\bibitem [{\citenamefont {Coleman}\ \emph {et~al.}(2011)\citenamefont
  {Coleman}, \citenamefont {Lotya}, \citenamefont {O’Neill}, \citenamefont
  {Bergin}, \citenamefont {King}, \citenamefont {Khan}, \citenamefont {Young},
  \citenamefont {Gaucher}, \citenamefont {De}, \citenamefont {Smith},
  \citenamefont {Shvets}, \citenamefont {Arora}, \citenamefont {Stanton},
  \citenamefont {Kim}, \citenamefont {Lee}, \citenamefont {Kim}, \citenamefont
  {Duesberg}, \citenamefont {Hallam}, \citenamefont {Boland}, \citenamefont
  {Wang}, \citenamefont {Donegan}, \citenamefont {Grunlan}, \citenamefont
  {Moriarty}, \citenamefont {Shmeliov}, \citenamefont {Nicholls}, \citenamefont
  {Perkins}, \citenamefont {Grieveson}, \citenamefont {Theuwissen},
  \citenamefont {McComb}, \citenamefont {Nellist},\ and\ \citenamefont
  {Nicolosi}}]{coleman2011}%
  \BibitemOpen
  \bibfield  {author} {\bibinfo {author} {\bibfnamefont {J.~N.}\ \bibnamefont
  {Coleman}}, \bibinfo {author} {\bibfnamefont {M.}~\bibnamefont {Lotya}},
  \bibinfo {author} {\bibfnamefont {A.}~\bibnamefont {O’Neill}}, \bibinfo
  {author} {\bibfnamefont {S.~D.}\ \bibnamefont {Bergin}}, \bibinfo {author}
  {\bibfnamefont {P.~J.}\ \bibnamefont {King}}, \bibinfo {author}
  {\bibfnamefont {U.}~\bibnamefont {Khan}}, \bibinfo {author} {\bibfnamefont
  {K.}~\bibnamefont {Young}}, \bibinfo {author} {\bibfnamefont
  {A.}~\bibnamefont {Gaucher}}, \bibinfo {author} {\bibfnamefont
  {S.}~\bibnamefont {De}}, \bibinfo {author} {\bibfnamefont {R.~J.}\
  \bibnamefont {Smith}}, \bibinfo {author} {\bibfnamefont {I.~V.}\ \bibnamefont
  {Shvets}}, \bibinfo {author} {\bibfnamefont {S.~K.}\ \bibnamefont {Arora}},
  \bibinfo {author} {\bibfnamefont {G.}~\bibnamefont {Stanton}}, \bibinfo
  {author} {\bibfnamefont {H.-Y.}\ \bibnamefont {Kim}}, \bibinfo {author}
  {\bibfnamefont {K.}~\bibnamefont {Lee}}, \bibinfo {author} {\bibfnamefont
  {G.~T.}\ \bibnamefont {Kim}}, \bibinfo {author} {\bibfnamefont {G.~S.}\
  \bibnamefont {Duesberg}}, \bibinfo {author} {\bibfnamefont {T.}~\bibnamefont
  {Hallam}}, \bibinfo {author} {\bibfnamefont {J.~J.}\ \bibnamefont {Boland}},
  \bibinfo {author} {\bibfnamefont {J.~J.}\ \bibnamefont {Wang}}, \bibinfo
  {author} {\bibfnamefont {J.~F.}\ \bibnamefont {Donegan}}, \bibinfo {author}
  {\bibfnamefont {J.~C.}\ \bibnamefont {Grunlan}}, \bibinfo {author}
  {\bibfnamefont {G.}~\bibnamefont {Moriarty}}, \bibinfo {author}
  {\bibfnamefont {A.}~\bibnamefont {Shmeliov}}, \bibinfo {author}
  {\bibfnamefont {R.~J.}\ \bibnamefont {Nicholls}}, \bibinfo {author}
  {\bibfnamefont {J.~M.}\ \bibnamefont {Perkins}}, \bibinfo {author}
  {\bibfnamefont {E.~M.}\ \bibnamefont {Grieveson}}, \bibinfo {author}
  {\bibfnamefont {K.}~\bibnamefont {Theuwissen}}, \bibinfo {author}
  {\bibfnamefont {D.~W.}\ \bibnamefont {McComb}}, \bibinfo {author}
  {\bibfnamefont {P.~D.}\ \bibnamefont {Nellist}}, \ and\ \bibinfo {author}
  {\bibfnamefont {V.}~\bibnamefont {Nicolosi}},\ }\href {\doibase
  10.1126/science.1194975} {\bibfield  {journal} {\bibinfo  {journal}
  {Science}\ }\textbf {\bibinfo {volume} {331}},\ \bibinfo {pages} {568}
  (\bibinfo {year} {2011})}\BibitemShut {NoStop}%
\bibitem [{\citenamefont {Novoselov}\ and\ \citenamefont
  {Neto}(2012)}]{novoselov2012}%
  \BibitemOpen
  \bibfield  {author} {\bibinfo {author} {\bibfnamefont {K.~S.}\ \bibnamefont
  {Novoselov}}\ and\ \bibinfo {author} {\bibfnamefont {A.~H.~C.}\ \bibnamefont
  {Neto}},\ }\href {http://stacks.iop.org/1402-4896/2012/i=T146/a=014006}
  {\bibfield  {journal} {\bibinfo  {journal} {Physica Scripta}\ }\textbf
  {\bibinfo {volume} {2012}},\ \bibinfo {pages} {014006} (\bibinfo {year}
  {2012})}\BibitemShut {NoStop}%
\bibitem [{\citenamefont {Mak}\ \emph {et~al.}(2010)\citenamefont {Mak},
  \citenamefont {Lee}, \citenamefont {Hone}, \citenamefont {Shan},\ and\
  \citenamefont {Heinz}}]{mak2010}%
  \BibitemOpen
  \bibfield  {author} {\bibinfo {author} {\bibfnamefont {K.~F.}\ \bibnamefont
  {Mak}}, \bibinfo {author} {\bibfnamefont {C.}~\bibnamefont {Lee}}, \bibinfo
  {author} {\bibfnamefont {J.}~\bibnamefont {Hone}}, \bibinfo {author}
  {\bibfnamefont {J.}~\bibnamefont {Shan}}, \ and\ \bibinfo {author}
  {\bibfnamefont {T.~F.}\ \bibnamefont {Heinz}},\ }\href {\doibase
  10.1103/PhysRevLett.105.136805} {\bibfield  {journal} {\bibinfo  {journal}
  {Phys. Rev. Lett.}\ }\textbf {\bibinfo {volume} {105}},\ \bibinfo {pages}
  {136805} (\bibinfo {year} {2010})}\BibitemShut {NoStop}%
\bibitem [{\citenamefont {Splendiani}\ \emph {et~al.}(2010)\citenamefont
  {Splendiani}, \citenamefont {Sun}, \citenamefont {Zhang}, \citenamefont {Li},
  \citenamefont {Kim}, \citenamefont {Chim}, \citenamefont {Galli},\ and\
  \citenamefont {Wang}}]{splendiani2010}%
  \BibitemOpen
  \bibfield  {author} {\bibinfo {author} {\bibfnamefont {A.}~\bibnamefont
  {Splendiani}}, \bibinfo {author} {\bibfnamefont {L.}~\bibnamefont {Sun}},
  \bibinfo {author} {\bibfnamefont {Y.}~\bibnamefont {Zhang}}, \bibinfo
  {author} {\bibfnamefont {T.}~\bibnamefont {Li}}, \bibinfo {author}
  {\bibfnamefont {J.}~\bibnamefont {Kim}}, \bibinfo {author} {\bibfnamefont
  {C.-Y.}\ \bibnamefont {Chim}}, \bibinfo {author} {\bibfnamefont
  {G.}~\bibnamefont {Galli}}, \ and\ \bibinfo {author} {\bibfnamefont
  {F.}~\bibnamefont {Wang}},\ }\href {\doibase 10.1021/nl903868w} {\bibfield
  {journal} {\bibinfo  {journal} {Nano Letters}\ }\textbf {\bibinfo {volume}
  {10}},\ \bibinfo {pages} {1271} (\bibinfo {year} {2010})}\BibitemShut
  {NoStop}%
\bibitem [{\citenamefont {Mak}\ \emph {et~al.}(2012{\natexlab{a}})\citenamefont
  {Mak}, \citenamefont {He}, \citenamefont {Shan},\ and\ \citenamefont
  {Heinz}}]{mak2012}%
  \BibitemOpen
  \bibfield  {author} {\bibinfo {author} {\bibfnamefont {K.~F.}\ \bibnamefont
  {Mak}}, \bibinfo {author} {\bibfnamefont {K.}~\bibnamefont {He}}, \bibinfo
  {author} {\bibfnamefont {J.}~\bibnamefont {Shan}}, \ and\ \bibinfo {author}
  {\bibfnamefont {T.~F.}\ \bibnamefont {Heinz}},\ }\href {\doibase
  doi:10.1038/nnano.2012.96} {\bibfield  {journal} {\bibinfo  {journal} {Nature
  Nanotechnology}\ }\textbf {\bibinfo {volume} {7}},\ \bibinfo {pages} {494}
  (\bibinfo {year} {2012}{\natexlab{a}})}\BibitemShut {NoStop}%
\bibitem [{\citenamefont {Zeng}\ \emph {et~al.}(2012)\citenamefont {Zeng},
  \citenamefont {Dai}, \citenamefont {Yao}, \citenamefont {Xiao},\ and\
  \citenamefont {Cui}}]{zeng2012}%
  \BibitemOpen
  \bibfield  {author} {\bibinfo {author} {\bibfnamefont {H.}~\bibnamefont
  {Zeng}}, \bibinfo {author} {\bibfnamefont {J.}~\bibnamefont {Dai}}, \bibinfo
  {author} {\bibfnamefont {W.}~\bibnamefont {Yao}}, \bibinfo {author}
  {\bibfnamefont {D.}~\bibnamefont {Xiao}}, \ and\ \bibinfo {author}
  {\bibfnamefont {X.}~\bibnamefont {Cui}},\ }\href {\doibase
  doi:10.1038/nnano.2012.95} {\bibfield  {journal} {\bibinfo  {journal} {Nature
  Nanotechnology}\ }\textbf {\bibinfo {volume} {7}},\ \bibinfo {pages} {490}
  (\bibinfo {year} {2012})}\BibitemShut {NoStop}%
\bibitem [{\citenamefont {Cao}\ \emph {et~al.}(2012)\citenamefont {Cao},
  \citenamefont {Wang}, \citenamefont {Han}, \citenamefont {Ye}, \citenamefont
  {Zhu}, \citenamefont {Shi}, \citenamefont {Niu}, \citenamefont {Tan},
  \citenamefont {Wang}, \citenamefont {Liu},\ and\ \citenamefont
  {Feng}}]{cao2012}%
  \BibitemOpen
  \bibfield  {author} {\bibinfo {author} {\bibfnamefont {T.}~\bibnamefont
  {Cao}}, \bibinfo {author} {\bibfnamefont {G.}~\bibnamefont {Wang}}, \bibinfo
  {author} {\bibfnamefont {W.}~\bibnamefont {Han}}, \bibinfo {author}
  {\bibfnamefont {H.}~\bibnamefont {Ye}}, \bibinfo {author} {\bibfnamefont
  {C.}~\bibnamefont {Zhu}}, \bibinfo {author} {\bibfnamefont {J.}~\bibnamefont
  {Shi}}, \bibinfo {author} {\bibfnamefont {Q.}~\bibnamefont {Niu}}, \bibinfo
  {author} {\bibfnamefont {P.}~\bibnamefont {Tan}}, \bibinfo {author}
  {\bibfnamefont {E.}~\bibnamefont {Wang}}, \bibinfo {author} {\bibfnamefont
  {B.}~\bibnamefont {Liu}}, \ and\ \bibinfo {author} {\bibfnamefont
  {J.}~\bibnamefont {Feng}},\ }\href {\doibase 10.1038/ncomms1882} {\bibfield
  {journal} {\bibinfo  {journal} {Nature Communications}\ }\textbf {\bibinfo
  {volume} {3}},\ \bibinfo {pages} {887} (\bibinfo {year} {2012})}\BibitemShut
  {NoStop}%
\bibitem [{\citenamefont {Tongay}\ \emph {et~al.}(2012)\citenamefont {Tongay},
  \citenamefont {Zhou}, \citenamefont {Ataca}, \citenamefont {Lo},
  \citenamefont {Matthews}, \citenamefont {Li}, \citenamefont {Grossman},\ and\
  \citenamefont {Wu}}]{tongay2012}%
  \BibitemOpen
  \bibfield  {author} {\bibinfo {author} {\bibfnamefont {S.}~\bibnamefont
  {Tongay}}, \bibinfo {author} {\bibfnamefont {J.}~\bibnamefont {Zhou}},
  \bibinfo {author} {\bibfnamefont {C.}~\bibnamefont {Ataca}}, \bibinfo
  {author} {\bibfnamefont {K.}~\bibnamefont {Lo}}, \bibinfo {author}
  {\bibfnamefont {T.~S.}\ \bibnamefont {Matthews}}, \bibinfo {author}
  {\bibfnamefont {J.}~\bibnamefont {Li}}, \bibinfo {author} {\bibfnamefont
  {J.~C.}\ \bibnamefont {Grossman}}, \ and\ \bibinfo {author} {\bibfnamefont
  {J.}~\bibnamefont {Wu}},\ }\href {\doibase 10.1021/nl302584w} {\bibfield
  {journal} {\bibinfo  {journal} {Nano Letters}\ }\textbf {\bibinfo {volume}
  {12}},\ \bibinfo {pages} {5576–5580} (\bibinfo {year} {2012})}\BibitemShut
  {NoStop}%
\bibitem [{\citenamefont {Wang}\ \emph {et~al.}(2012)\citenamefont {Wang},
  \citenamefont {Kalantar-Zadeh}, \citenamefont {Kis}, \citenamefont
  {Coleman},\ and\ \citenamefont {Strano}}]{wang2012}%
  \BibitemOpen
  \bibfield  {author} {\bibinfo {author} {\bibfnamefont {Q.~H.}\ \bibnamefont
  {Wang}}, \bibinfo {author} {\bibfnamefont {K.}~\bibnamefont
  {Kalantar-Zadeh}}, \bibinfo {author} {\bibfnamefont {A.}~\bibnamefont {Kis}},
  \bibinfo {author} {\bibfnamefont {J.~N.}\ \bibnamefont {Coleman}}, \ and\
  \bibinfo {author} {\bibfnamefont {M.~S.}\ \bibnamefont {Strano}},\ }\href
  {\doibase 10.1038/nnano.2012.193} {\bibfield  {journal} {\bibinfo  {journal}
  {Nature Nanotechnology}\ }\textbf {\bibinfo {volume} {7}},\ \bibinfo {pages}
  {699} (\bibinfo {year} {2012})}\BibitemShut {NoStop}%
\bibitem [{\citenamefont {Mak}\ \emph {et~al.}(2012{\natexlab{b}})\citenamefont
  {Mak}, \citenamefont {He}, \citenamefont {Lee}, \citenamefont {Lee},
  \citenamefont {Hone}, \citenamefont {Heinz},\ and\ \citenamefont
  {Shan}}]{mak2012b}%
  \BibitemOpen
  \bibfield  {author} {\bibinfo {author} {\bibfnamefont {K.~F.}\ \bibnamefont
  {Mak}}, \bibinfo {author} {\bibfnamefont {K.}~\bibnamefont {He}}, \bibinfo
  {author} {\bibfnamefont {C.}~\bibnamefont {Lee}}, \bibinfo {author}
  {\bibfnamefont {G.~H.}\ \bibnamefont {Lee}}, \bibinfo {author} {\bibfnamefont
  {J.}~\bibnamefont {Hone}}, \bibinfo {author} {\bibfnamefont {T.~F.}\
  \bibnamefont {Heinz}}, \ and\ \bibinfo {author} {\bibfnamefont
  {J.}~\bibnamefont {Shan}},\ }\href {\doibase doi: 10.1038/nmat3505}
  {\bibfield  {journal} {\bibinfo  {journal} {Nature Materials}\ } (\bibinfo
  {year} {2012}{\natexlab{b}}),\ doi: 10.1038/nmat3505}\BibitemShut {NoStop}%
\bibitem [{\citenamefont {Radisavljevic}\ \emph {et~al.}(2011)\citenamefont
  {Radisavljevic}, \citenamefont {Radenovic}, \citenamefont {Brivio},\ and\
  \citenamefont {Kis}}]{radisavljevic2011}%
  \BibitemOpen
  \bibfield  {author} {\bibinfo {author} {\bibfnamefont {B.}~\bibnamefont
  {Radisavljevic}}, \bibinfo {author} {\bibfnamefont {A.}~\bibnamefont
  {Radenovic}}, \bibinfo {author} {\bibfnamefont {J.}~\bibnamefont {Brivio}}, \
  and\ \bibinfo {author} {\bibfnamefont {V.~G.~A.}\ \bibnamefont {Kis}},\
  }\href {\doibase doi:10.1038/nnano.2010.279} {\bibfield  {journal} {\bibinfo
  {journal} {Nature Nanotechnology}\ }\textbf {\bibinfo {volume} {6}},\
  \bibinfo {pages} {147} (\bibinfo {year} {2011})}\BibitemShut {NoStop}%
\bibitem [{\citenamefont {Yin}\ \emph {et~al.}(2012)\citenamefont {Yin},
  \citenamefont {Li}, \citenamefont {Li}, \citenamefont {Jiang}, \citenamefont
  {Shi}, \citenamefont {Sun}, \citenamefont {Lu}, \citenamefont {Zhang},
  \citenamefont {Chen},\ and\ \citenamefont {Zhang}}]{yin2012}%
  \BibitemOpen
  \bibfield  {author} {\bibinfo {author} {\bibfnamefont {Z.}~\bibnamefont
  {Yin}}, \bibinfo {author} {\bibfnamefont {H.}~\bibnamefont {Li}}, \bibinfo
  {author} {\bibfnamefont {H.}~\bibnamefont {Li}}, \bibinfo {author}
  {\bibfnamefont {L.}~\bibnamefont {Jiang}}, \bibinfo {author} {\bibfnamefont
  {Y.}~\bibnamefont {Shi}}, \bibinfo {author} {\bibfnamefont {Y.}~\bibnamefont
  {Sun}}, \bibinfo {author} {\bibfnamefont {G.}~\bibnamefont {Lu}}, \bibinfo
  {author} {\bibfnamefont {Q.}~\bibnamefont {Zhang}}, \bibinfo {author}
  {\bibfnamefont {X.}~\bibnamefont {Chen}}, \ and\ \bibinfo {author}
  {\bibfnamefont {H.}~\bibnamefont {Zhang}},\ }\href {\doibase
  10.1021/nn2024557} {\bibfield  {journal} {\bibinfo  {journal} {ACS Nano}\
  }\textbf {\bibinfo {volume} {6}},\ \bibinfo {pages} {74} (\bibinfo {year}
  {2012})}\BibitemShut {NoStop}%
\bibitem [{\citenamefont {Xiao}\ \emph {et~al.}(2012)\citenamefont {Xiao},
  \citenamefont {Liu}, \citenamefont {Feng}, \citenamefont {Xu},\ and\
  \citenamefont {Yao}}]{xiao2012}%
  \BibitemOpen
  \bibfield  {author} {\bibinfo {author} {\bibfnamefont {D.}~\bibnamefont
  {Xiao}}, \bibinfo {author} {\bibfnamefont {G.-B.}\ \bibnamefont {Liu}},
  \bibinfo {author} {\bibfnamefont {W.}~\bibnamefont {Feng}}, \bibinfo {author}
  {\bibfnamefont {X.}~\bibnamefont {Xu}}, \ and\ \bibinfo {author}
  {\bibfnamefont {W.}~\bibnamefont {Yao}},\ }\href {\doibase
  10.1103/PhysRevLett.108.196802} {\bibfield  {journal} {\bibinfo  {journal}
  {Phys. Rev. Lett.}\ }\textbf {\bibinfo {volume} {108}},\ \bibinfo {pages}
  {196802} (\bibinfo {year} {2012})}\BibitemShut {NoStop}%
\bibitem [{\citenamefont {Boyd}(2008)}]{boyd}%
  \BibitemOpen
  \bibfield  {author} {\bibinfo {author} {\bibfnamefont {R.}~\bibnamefont
  {Boyd}},\ }\href@noop {} {\emph {\bibinfo {title} {Nonlinear optics}}}\
  (\bibinfo  {publisher} {Academic Press, London},\ \bibinfo {year}
  {2008})\BibitemShut {NoStop}%
\bibitem [{\citenamefont {Shen}(2003)}]{shen}%
  \BibitemOpen
  \bibfield  {author} {\bibinfo {author} {\bibfnamefont {Y.~R.}\ \bibnamefont
  {Shen}},\ }\href@noop {} {\emph {\bibinfo {title} {The Principles of
  Nonlinear Optics}}}\ (\bibinfo  {publisher} {Wiley Interscience, New York},\
  \bibinfo {year} {2003})\BibitemShut {NoStop}%
\bibitem [{\citenamefont {Lee}\ \emph {et~al.}(2010)\citenamefont {Lee},
  \citenamefont {Yan}, \citenamefont {Brus}, \citenamefont {Heinz},
  \citenamefont {Hone},\ and\ \citenamefont {Ryu}}]{lee2010}%
  \BibitemOpen
  \bibfield  {author} {\bibinfo {author} {\bibfnamefont {C.}~\bibnamefont
  {Lee}}, \bibinfo {author} {\bibfnamefont {H.}~\bibnamefont {Yan}}, \bibinfo
  {author} {\bibfnamefont {L.~E.}\ \bibnamefont {Brus}}, \bibinfo {author}
  {\bibfnamefont {T.~F.}\ \bibnamefont {Heinz}}, \bibinfo {author}
  {\bibfnamefont {J.}~\bibnamefont {Hone}}, \ and\ \bibinfo {author}
  {\bibfnamefont {S.}~\bibnamefont {Ryu}},\ }\href {\doibase 10.1021/nn1003937}
  {\bibfield  {journal} {\bibinfo  {journal} {ACS Nano}\ }\textbf {\bibinfo
  {volume} {4}},\ \bibinfo {pages} {2695} (\bibinfo {year} {2010})}\BibitemShut
  {NoStop}%
\bibitem [{\citenamefont {Benameur}\ \emph {et~al.}(2011)\citenamefont
  {Benameur}, \citenamefont {Radisavljevic}, \citenamefont {Héron},
  \citenamefont {Sahoo}, \citenamefont {Berger},\ and\ \citenamefont
  {Kis}}]{benameur2011}%
  \BibitemOpen
  \bibfield  {author} {\bibinfo {author} {\bibfnamefont {M.~M.}\ \bibnamefont
  {Benameur}}, \bibinfo {author} {\bibfnamefont {B.}~\bibnamefont
  {Radisavljevic}}, \bibinfo {author} {\bibfnamefont {J.~S.}\ \bibnamefont
  {Héron}}, \bibinfo {author} {\bibfnamefont {S.}~\bibnamefont {Sahoo}},
  \bibinfo {author} {\bibfnamefont {H.}~\bibnamefont {Berger}}, \ and\ \bibinfo
  {author} {\bibfnamefont {A.}~\bibnamefont {Kis}},\ }\href
  {http://stacks.iop.org/0957-4484/22/i=12/a=125706} {\bibfield  {journal}
  {\bibinfo  {journal} {Nanotechnology}\ }\textbf {\bibinfo {volume} {22}},\
  \bibinfo {pages} {125706} (\bibinfo {year} {2011})}\BibitemShut {NoStop}%
\bibitem [{\citenamefont {Tom}\ \emph {et~al.}(1983)\citenamefont {Tom},
  \citenamefont {Heinz},\ and\ \citenamefont {Shen}}]{tom1983}%
  \BibitemOpen
  \bibfield  {author} {\bibinfo {author} {\bibfnamefont {H.~W.~K.}\
  \bibnamefont {Tom}}, \bibinfo {author} {\bibfnamefont {T.~F.}\ \bibnamefont
  {Heinz}}, \ and\ \bibinfo {author} {\bibfnamefont {Y.~R.}\ \bibnamefont
  {Shen}},\ }\href {\doibase 10.1103/PhysRevLett.51.1983} {\bibfield  {journal}
  {\bibinfo  {journal} {Phys. Rev. Lett.}\ }\textbf {\bibinfo {volume} {51}},\
  \bibinfo {pages} {1983} (\bibinfo {year} {1983})}\BibitemShut {NoStop}%
\bibitem [{\citenamefont {Shen}(1989)}]{shen89}%
  \BibitemOpen
  \bibfield  {author} {\bibinfo {author} {\bibfnamefont {Y.~R.}\ \bibnamefont
  {Shen}},\ }\href {\doibase 10.1146/annurev.pc.40.100189.001551} {\bibfield
  {journal} {\bibinfo  {journal} {Annual Review of Physical Chemistry}\
  }\textbf {\bibinfo {volume} {40}},\ \bibinfo {pages} {327} (\bibinfo {year}
  {1989})}\BibitemShut {NoStop}%
\bibitem [{\citenamefont {Mattheiss}(1973)}]{mattheiss1973}%
  \BibitemOpen
  \bibfield  {author} {\bibinfo {author} {\bibfnamefont {L.~F.}\ \bibnamefont
  {Mattheiss}},\ }\href {\doibase 10.1103/PhysRevB.8.3719} {\bibfield
  {journal} {\bibinfo  {journal} {Phys. Rev. B}\ }\textbf {\bibinfo {volume}
  {8}},\ \bibinfo {pages} {3719} (\bibinfo {year} {1973})}\BibitemShut
  {NoStop}%
\bibitem [{\citenamefont {Dresselhaus}\ \emph {et~al.}(2008)\citenamefont
  {Dresselhaus}, \citenamefont {Dresselhaus},\ and\ \citenamefont
  {Jorio}}]{dresselhaus}%
  \BibitemOpen
  \bibfield  {author} {\bibinfo {author} {\bibfnamefont {M.~S.}\ \bibnamefont
  {Dresselhaus}}, \bibinfo {author} {\bibfnamefont {G.}~\bibnamefont
  {Dresselhaus}}, \ and\ \bibinfo {author} {\bibfnamefont {A.}~\bibnamefont
  {Jorio}},\ }\href@noop {} {\emph {\bibinfo {title} {Group Theory Applications
  to the Physics of Condensed Matter}}}\ (\bibinfo  {publisher}
  {Springer-Verlag, Berlin},\ \bibinfo {year} {2008})\BibitemShut {NoStop}%
\bibitem [{\citenamefont {Malard}\ \emph {et~al.}(2009)\citenamefont {Malard},
  \citenamefont {Guimar\~aes}, \citenamefont {Mafra}, \citenamefont {Mazzoni},\
  and\ \citenamefont {Jorio}}]{malard2009}%
  \BibitemOpen
  \bibfield  {author} {\bibinfo {author} {\bibfnamefont {L.~M.}\ \bibnamefont
  {Malard}}, \bibinfo {author} {\bibfnamefont {M.~H.~D.}\ \bibnamefont
  {Guimar\~aes}}, \bibinfo {author} {\bibfnamefont {D.~L.}\ \bibnamefont
  {Mafra}}, \bibinfo {author} {\bibfnamefont {M.~S.~C.}\ \bibnamefont
  {Mazzoni}}, \ and\ \bibinfo {author} {\bibfnamefont {A.}~\bibnamefont
  {Jorio}},\ }\href {\doibase 10.1103/PhysRevB.79.125426} {\bibfield  {journal}
  {\bibinfo  {journal} {Phys. Rev. B}\ }\textbf {\bibinfo {volume} {79}},\
  \bibinfo {pages} {125426} (\bibinfo {year} {2009})}\BibitemShut {NoStop}%
\bibitem [{\citenamefont {Son}\ \emph {et~al.}(2006)\citenamefont {Son},
  \citenamefont {Cohen},\ and\ \citenamefont {Louie}}]{louie2006}%
  \BibitemOpen
  \bibfield  {author} {\bibinfo {author} {\bibfnamefont {Y.-W.}\ \bibnamefont
  {Son}}, \bibinfo {author} {\bibfnamefont {M.~L.}\ \bibnamefont {Cohen}}, \
  and\ \bibinfo {author} {\bibfnamefont {S.~G.}\ \bibnamefont {Louie}},\ }\href
  {\doibase 10.1038/nature05180} {\bibfield  {journal} {\bibinfo  {journal}
  {Nature}\ }\textbf {\bibinfo {volume} {444}},\ \bibinfo {pages} {347}
  (\bibinfo {year} {2006})}\BibitemShut {NoStop}%
\bibitem [{\citenamefont {Giovannetti}\ \emph {et~al.}(2007)\citenamefont
  {Giovannetti}, \citenamefont {Khomyakov}, \citenamefont {Brocks},
  \citenamefont {Kelly},\ and\ \citenamefont {van~den Brink}}]{jeroen2007}%
  \BibitemOpen
  \bibfield  {author} {\bibinfo {author} {\bibfnamefont {G.}~\bibnamefont
  {Giovannetti}}, \bibinfo {author} {\bibfnamefont {P.~A.}\ \bibnamefont
  {Khomyakov}}, \bibinfo {author} {\bibfnamefont {G.}~\bibnamefont {Brocks}},
  \bibinfo {author} {\bibfnamefont {P.~J.}\ \bibnamefont {Kelly}}, \ and\
  \bibinfo {author} {\bibfnamefont {J.}~\bibnamefont {van~den Brink}},\ }\href
  {\doibase 10.1103/PhysRevB.76.073103} {\bibfield  {journal} {\bibinfo
  {journal} {Phys. Rev. B}\ }\textbf {\bibinfo {volume} {76}},\ \bibinfo
  {pages} {073103} (\bibinfo {year} {2007})}\BibitemShut {NoStop}%
\bibitem [{\citenamefont {Yankowitz}\ \emph {et~al.}(2012)\citenamefont
  {Yankowitz}, \citenamefont {Xue}, \citenamefont {Cormode}, \citenamefont
  {Sanchez-Yamagishi}, \citenamefont {Watanabe}, \citenamefont {Taniguchi},
  \citenamefont {Jarillo-Herrero}, \citenamefont {Jacquod},\ and\ \citenamefont
  {LeRoy}}]{leroy2012}%
  \BibitemOpen
  \bibfield  {author} {\bibinfo {author} {\bibfnamefont {M.}~\bibnamefont
  {Yankowitz}}, \bibinfo {author} {\bibfnamefont {J.}~\bibnamefont {Xue}},
  \bibinfo {author} {\bibfnamefont {D.}~\bibnamefont {Cormode}}, \bibinfo
  {author} {\bibfnamefont {J.~D.}\ \bibnamefont {Sanchez-Yamagishi}}, \bibinfo
  {author} {\bibfnamefont {K.}~\bibnamefont {Watanabe}}, \bibinfo {author}
  {\bibfnamefont {T.}~\bibnamefont {Taniguchi}}, \bibinfo {author}
  {\bibfnamefont {P.}~\bibnamefont {Jarillo-Herrero}}, \bibinfo {author}
  {\bibfnamefont {P.}~\bibnamefont {Jacquod}}, \ and\ \bibinfo {author}
  {\bibfnamefont {B.~J.}\ \bibnamefont {LeRoy}},\ }\href {\doibase
  10.1038/nphys2272} {\bibfield  {journal} {\bibinfo  {journal} {Nature
  Physics}\ }\textbf {\bibinfo {volume} {8}},\ \bibinfo {pages} {382} (\bibinfo
  {year} {2012})}\BibitemShut {NoStop}%
\bibitem [{\citenamefont {Bloembergen}\ and\ \citenamefont
  {Pershan}(1962)}]{bloembergen1962}%
  \BibitemOpen
  \bibfield  {author} {\bibinfo {author} {\bibfnamefont {N.}~\bibnamefont
  {Bloembergen}}\ and\ \bibinfo {author} {\bibfnamefont {P.~S.}\ \bibnamefont
  {Pershan}},\ }\href {\doibase 10.1103/PhysRev.128.606} {\bibfield  {journal}
  {\bibinfo  {journal} {Phys. Rev.}\ }\textbf {\bibinfo {volume} {128}},\
  \bibinfo {pages} {606} (\bibinfo {year} {1962})}\BibitemShut {NoStop}%
\end{thebibliography}
%\bibliographystyle{apsrev4-1}

%merlin.mbs apsrev4-1.bst 2010-07-25 4.21a (PWD, AO, DPC) hacked
%Control: key (0)
%Control: author (72) initials jnrlst
%Control: editor formatted (1) identically to author
%Control: production of article title (-1) disabled
%Control: page (0) single
%Control: year (1) truncated
%Control: production of eprint (0) enabled
%

\end{document}